\documentclass[final,5p,times,twocolumn]{elsarticle}

\usepackage{graphics}
 \usepackage{graphicx}
 \usepackage{epstopdf}
 \usepackage{epsfig}
\usepackage{amssymb}
\usepackage{amsmath}
\usepackage{lineno}
\usepackage{tikz}
\usepackage{lipsum}

\newcommand\copyrighttext{%
  \footnotesize Published in Nuclear Instruments and Methods in Physics Research A 740 (2014) 173-179, DOI:10.1016/j.nima.2013.11.006}
\newcommand\copyrightnotice{%
\begin{tikzpicture}[remember picture,overlay]
\node[anchor=south,yshift=10pt] at (current page.south) {\fbox{\parbox{\dimexpr\textwidth-\fboxsep-\fboxrule\relax}{\copyrighttext}}};
\end{tikzpicture}%
}

\journal{Nuclear Instruments and Methods in Physics Research A }

\begin{document}

%\copyrightnotice
%\lipsum[1-1]

\begin{frontmatter}



\title{Collider design issues based on proton-driven plasma wakefield acceleration}


\author{G. Xia$^{a,b}$, O. Mete$^{a,b}$, A. Aimidula$^{b,c}$, C. Welsch$^{b,c}$, S. Chattopadhyay$^{a,b,c}$, S. Mandry$^{d}$, M. Wing$^{d,e}$}

\address[a]{School of Physics and Astronomy, University of Manchester, Manchester, United Kingdom}
\address[b]{The Cockcroft Institute, Sci-Tech Daresbury, Daresbury, Warrington, United Kingdom}
\address[c]{The University of Liverpool, Liverpool, United Kingdom}
\address[d]{Department of Physics and Astronomy, University College London, London, United Kingdom}
\address[e]{Deutsche Elektronen-Synchrotron DESY, Hamburg, Germany}

\begin{abstract}
Recent simulations have shown that a high-energy proton bunch can excite strong plasma wakefields and accelerate a bunch of electrons to the energy frontier in a single stage of acceleration. It therefore paves the way towards a compact future collider design using the proton beams from existing high-energy proton machines, e.g. Tevatron or the LHC. This paper addresses some key issues in designing a compact electron-positron linear collider and an electron-proton collider based on existing CERN accelerator infrastructure.
\end{abstract}

\begin{keyword}
PDPWA; Colliders; Self-modulation instability; Dephasing

\end{keyword}

\end{frontmatter}
\copyrightnotice

% \linenumbers

\section{Introduction}
\label{}
%\begin{figure}[h!] 
 %\centering
%\caption[my image]{Livingstone Chart (Updated).}
%\end{figure}

With the recent discovery of the Higgs boson at the Large Hadron Collider (LHC) at CERN \cite{atlas,cms}, the high energy physics community anticipated the construction of a dedicated Higgs factory, which may be an electron-positron ($e^+e^-$) linear collider for the precise measurement of the properties of the Higgs particle, e.g., its mass, spin, couplings with other particles and self-couplings, etc. However, current $e^+e^-$ linear collider designs at the energy frontier (TeV, or $10^{12}$ electronvolts) such as the International Linear Collider (ILC) and Compact Linear Collider (CLIC) extends over 30 km and costs over multi-billion dollars. The sizes of these machines are heavily dependent on the length of the RF linac, which is subject to a maximum material breakdown field (of $\sim$150$\,$MeV/m) and is the main cost driver for next generation linear colliders. The obvious question is: can we make the future machine more compact and cost effective?

In addition, the possibility of a lepton-hadron (e.g. $ep$) collider at CERN has been of interest since the initial proposal of the LHC. It has long been known that lepton-hadron collisions play an important role in the exploration of the fundamental structure of matter. For example, the quark-parton model originated from investigation of electron-nucleon scattering. The current proposed LHeC design employs the LHC beam colliding with the electron beam from a newly designed energy recovery linac (ERL) based ring or from a linac \cite{Bruening}. However, this design is expensive, e.g. the ring based design needs about 9$\,$km tunnel and a 19$\,$km bending arcs. The electron beam power is greater than 100$\,$MW and the project is not listed as the high priority for the recently updated European strategy for particle physics \cite{strategy}.

The development of plasma accelerators has achieved tremendous progress in the last decade. Laser wakefield accelerators (LWFAs) can routinely produce $\sim$GeV electron beams of percentage energy spread with only a few centimeter plasma cell and the accelerating gradient ($\sim$100$\,$GeV/m) is more than three orders of magnitude higher than the fields in conventional RF based structures \cite{Leemans}. Charged particle beam driven plasma wakefield acceleration (PWFA) has successfully demonstrated energy doubling from 42 to 85$\,$GeV of the electron beam from the Stanford Linear Collider (SLC) within an 85$\,$cm plasma cell \cite{Muggli1}. These significant breakthroughs have shown great promise to make a future machine more compact and cheaper. Based on these LWFA/PWFA schemes, a future energy frontier linear collider will consist of multi-stages, on the order of 100/50, to reach the TeV energy scale with each stage yielding energy gains of $\sim$10/20$\,$GeV. It should be noted that the multi-stage scheme introduces new challenges such as tight synchronization and alignment requirements of the drive and witness bunches and of each accelerator module (plasma cell). Staging also means a gradient dilution due to long distances required between each accelerator module for coupling new drive bunches and to capture and refocus the very low beta function witness bunches \cite{Muggli2}.

Proton driven plasma wakefield acceleration (PDPWA) has been recently proposed as a means to accelerate a bunch of electrons to the energy frontier in a single stage of acceleration \cite{Caldwell1, Lotov}. The advantages of using the proton beam as driver compared to other drive beams such as electron beams and laser beams lie in the facts of the availability of high-energy proton beams and the extremely high energies stored in current proton beams. For instance, the energy stored at a TeV LHC-like proton bunch is in general more than two orders of magnitude higher than that of the nowadaysÕ maximum energies of electron bunches or a laser pulse. Particle-in-cell (PIC) simulations have shown that a 1$\,$TeV LHC-like proton bunch, if compressed longitudinally to 100 $\mu$m, may become an ideal drive beam and can excite a plasma wakefield with an average accelerating field of $\sim$2$\,$GeV/m. Surfing on the right phase, a bunch of electrons can sample the plasma wakefields and gain energies up to 600$\,$GeV in a single passage of a 500$\,$m plasma [8]. Although the peak gradient is modest compared to LWFA/PWFA schemes, it is very similar to the average gradient of a PWFA based collider and is reached at relatively low plasma density, i.e. in the range of $10^{14}-10^{15}\,cm^{-3}$. This relatively low plasma density leads to a relatively large accelerating structure, which can potentially relax the temporal and spatial alignment tolerances, as well as the witness beam parameters. If this scheme can be demonstrated, it will point to a new way for a compact TeV collider design based on existing TeV proton machines, e.g. the CERN accelerator complex. Compared to LWFA/PWFA based collider designs, this will greatly reduce the stringent requirement on the alignment and synchronization of the multi-stage accelerator modules. 

However, one hurdle in the above scheme is the proton bunch compression. Bunch compression via a magnetic chicane is a widely used method to compress the electron bunch to sub-millimetre scale. However, it is non-trivial to adopt this idea while still keeping the bunch charge constant.  It turns out that a large amount of RF power is needed to provide the energy chirp along the bunch and large dipole magnets are required to offer the energy-path correlation. Simulation shows that 4$\,$km of RF cavities are required to achieve this task \cite{Xia1}. This does not seem practical.

\section{Self-modulation of a long proton bunch}
\label{}
It has long been known that a long laser pulse can be modulated by a high-density plasma. This so-called self-modulated laser wakefield acceleration (SM-LWFA) has sustained the large wakefield amplitude of 100$\,$GeV/m \cite{Esarey}. In this scenario, the SM process occurs due to forward Raman scattering, i.e., the laser light scatters on the noise at the plasma period, which results in a wave shift by the plasma frequency. The two waves then beat together to drive the plasma wave. Eventually the long pulse is split into many ultra-short slices with a length of half of the plasma wavelength each separated by a plasma wavelength (note that the plasma wavelength is inversely proportional to the square root of the plasma density). Similarly, when a long proton bunch enters into a plasma, protons at the bunch head excite plasma wakefields. The transverse plasma wakefields can then focus and defocus the body of the driver bunch. In the case of a drive bunch much longer than the plasma wavelength, the bunch is subject to focusing and defocusing forces along the whole beam. The overall effect is that the long beam is modulated by the wakefields it produces. The resulting bunches have a slice length of half of the plasma wavelength, may contain a small portion of protons, with a distance of a plasma wavelength between each slice. Further investigation shows that it takes time for the modulation to occur, however, once the modulation starts and eventually saturates, these ultrashort proton bunch slices will excite plasma wakefields and the fields will add up coherently \cite{Kumar}. Recent simulations show that the maximum wakefield amplitude from a modulated proton bunch is comparable to that of a short bunch driver. For example, an LHC beam with a beam energy of 7$\,$TeV, a bunch intensity of $1.15 \times 10^{11}$ and an rms bunch length of 7.55$\,$cm can excite wakefields with maximum amplitude of $\sim$1.5$\,$GeV/m working in self-modulation regime at a plasma density of $3 \times 10^{15}\, cm^{-3}$. An externally injected electron bunch will be accelerated up to 6$\,$TeV after propagating through a 10$\,$km plasma \cite{Caldwell2}. This indicates that one may achieve a very high-energy electron beam by using today's long and high-energy proton bunch directly as drive beam, assuming we could make such a long plasma source for the experiment. Based on this self-modulated proton driven plasma wakefield acceleration scheme, future colliders, either an $e^+e^-$ collider or an $ep$ collider can be conceived. 

It should be noted that the recently proposed AWAKE experiment will test this PDPWA scheme by using the proton beam from CERN SPS \cite{Xia2}. In this experiment, a $450\,$GeV proton bunch enters a $\sim$10$\,$m plasma. The self-modulation of the long proton bunch will be experimentally observed and an externally injected witness electron beam with a beam energy of 10-20$\,$MeV will be accelerated by the plasma wakefields and gain an energy of about 2$\,$GeV. The AWAKE experiment at CERN will shed light on a future compact collider design from an experimental point of view \cite{Caldwell3}.
    
In this paper, we discuss some key issues in the design of a compact, multi-TeV collider considering an $e^+e^-$ linear collider and a high-energy $ep$ collider based on the PDPWA scheme. Two important parameters, i.e. center-of-mass energy and luminosity are discussed in section 3. Section 4 gives an example design of a 2$\,$TeV $e^+e^-$ linear collider based at the LHC tunnel. An $ep$ collider design consideration is introduced in section 5. Section 6 discusses some key issues, e.g. phase slippage, proton beam guiding in long plasma, electron scattering in plasma and positron acceleration in the collider design based on PDPWA scheme. Some other novel collider schemes based on PDPWA are also introduced in section 7.

\section{Center-of-mass energy and luminosity}
\label{}
There are two figures of merit for future colliders that characterize the interactions between two colliding beams, one of them is the center-of-mass (CoM) energy and the other is the luminosity. The CoM energy is determined by the physics process to be studied, while the luminosity gives the production rate for a particle of interest and therefore it determines the performance of a collider. For the electron-positron linear collider, the CoM energy is $E_{com}= 2E_b$, where $E_b$ is the energy per beam and we assume that the energies of electrons and positrons are exactly the same. And for an electron-proton collider, the CoM energy is given by,

\begin{equation}
E_{com}=2\sqrt{E_eE_p},
\label{eqn:Ecom}
\end{equation}
where $E_e$ and $E_p$ are the beam energy for electrons and protons, respectively.
     As the main design parameter for a linear collider, the next $e^+e^-$ collider is envisioned to be at the TeV scale with a luminosity of $10^{34}\,cm^{-2}s^{-1}$. For two Gaussian beams of electrons and positrons, the luminosity is given by,

\begin{equation}
\mathcal{L}=f\frac{N^+N^-}{4\pi\sigma^*_x\sigma^*_y},
\label{eqn:Luminosity1}
\end{equation}
where $f$ denotes the collision rate (frequency) of the beams, $N^+$ and $N^-$ are the bunch population for electrons and positrons ($N^+ = N^- = N$ if the bunch population for electrons and positrons are the same), $\sigma^*_x$ and $\sigma^*_y$ are the horizontal and vertical beam spot sizes at the interaction point (IP). 

The luminosity can be easily rewritten using the beam power, $P_b$:
                                                 
\begin{equation}
\mathcal{L}=\frac{P_bN}{4\pi E_b\sigma^*_x\sigma^*_y}.
\label{eqn:Luminosity2}
\end{equation}
                                                                                                           
From Eq. (3), one can conclude that for a fixed IP design, i.e. fixed beam energy and beam spot sizes at the IP, the luminosity is proportional to the average power of the beam and the number of particles per bunch. The average beam power for the current ILC of 500$\,$GeV CoM is about 10$\,$MW with a bunch population of $10^{10}$, a repetition rate of 10$\,$kHz and with each bunch energy of $\sim$kJ. In order to obtain the required luminosity of $10^{34}\, cm^{-2}s^{-1}$ in a TeV collider based on plasma wakefield acceleration scheme, the average power of the drive beam needs to be larger than 10$\,$MW since the coupling efficiency from the drive beam to witness beam is less than unity. The beam power of current high-energy proton machines, e.g., Tevatron or the LHC is much larger than this value. Table \ref{tab:beam_specs} gives the comparison of beam specifications between the current proton machines and the lepton machines. One can see clearly that the stored bunch energies for current hadron machines are about two to three orders of magnitude higher than that for the current most energetic electron machine FACET and the planned facilities such as ILC and CLIC. If the energy coupling efficiency is about percentage level from the drive beam (protons) to the witness beam (electrons) via plasma wakefields, one could expect to achieve the beam specifications for an $e^+e^-$ or an $ep$ collider.

\begin{table*}[t]
\caption{Parameters of particle beams in present and planned facilities.} % title of Table
\centering % used for centering table
{\small
\begin{tabular}{l c c c c c c} % centered columns (4 columns)
\hline\hline %inserts double horizontal lines
& FACET & ILC & CLIC & SPS & Tevatron & LHC \\ [0.5ex] % inserts table 
%heading
\hline % inserts single horizontal line
Beam energy (GeV) & 25	& 250 & 1500 & 450 & 1000 & 7000 \\ % inserting body of the table
Luminosity ($10^{34}$ $cm^{-2}s^{-1}$) & - & 2 & 6 & - & 0.04 & 1  \\
Bunch intensity	 ($10^{10}$) & 2.0 & 2.0 & 0.372 & 13 & 27 & 11.5  \\
Bunches per beam	& 1 & 2625 & 312 & 288 & 36 & 2808  \\
IP bunch length ($\mu$m) & 30 & 300 & 30 & 1.2E5 & 350 & 7.5E4\\
IP beam sizes $\sigma_x^*/\sigma_y^*$ (nm) &1.4E4/6.0E3 & 	474/5.9 & 40/1 & 200 & 3.3E4 & 1.6E4\\
Rep rate (Hz) &	 1 & 5 & 50 & - & 1 & 1\\
Stored bunch energy (kJ) & 0.08 & 0.8 & 0.89 & 9.4 & 43 & 129\\
Beam power (W) & 80 & 1.05E7 & 1.39E7 & - & 5.49E7 & 3.62E8 \\ [1ex] % [1ex] adds vertical space
\hline %inserts single line
\end{tabular} }
\label{table:nonlin} % is used to refer this table in the text
\label{tab:beam_specs}
\end{table*}

\section{An electron-positron linear collider}
\label{}

\begin{figure}[ht] 
\centering
\includegraphics[width=0.5\textwidth]{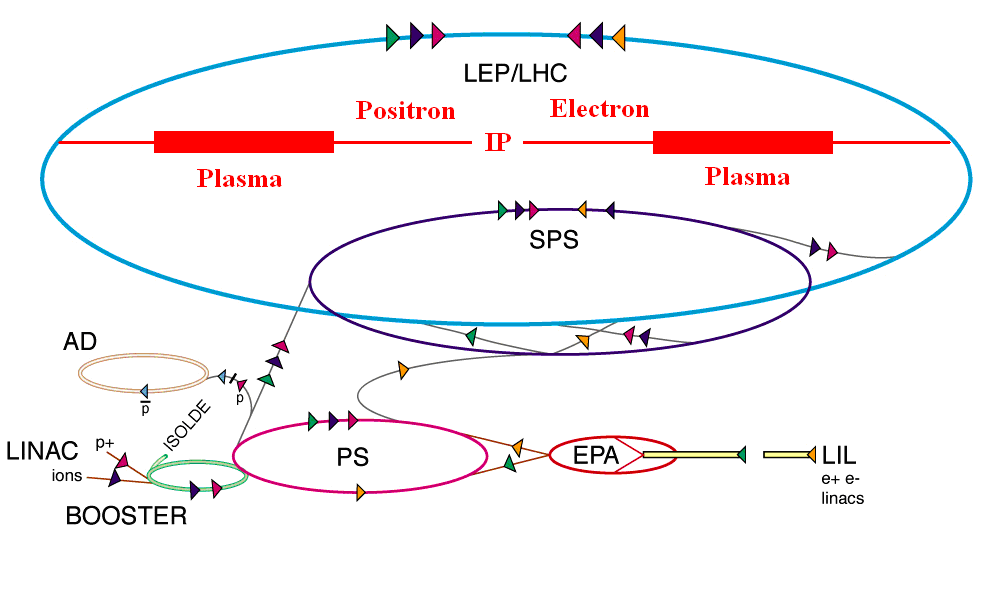}
\caption{Schematic layout of a 2 TeV CoM electron-positron linear collider based on a modulated proton-driven plasma wakefield acceleration.}
\label{fig:LHCILC}
\end{figure}

As we mentioned earlier, a modulated high-energy proton bunch can produce a high amplitude plasma wakefield and accelerate a trailing electron bunch to the energy frontier in a single stage of acceleration. Latest simulations show that a positron beam can also be accelerated in the wakefield from a modulated long proton bunch \cite{Vieire}. We can therefore conceive of a TeV $e^+e^-$ collider design based on this self-modulation PDPWA scheme. Simulation indicates that in this case the excited wakefield always shows a decay pattern. This is mainly due to the phase shift between the resulting bunch slices and the phase of the wakefields excited. To overcome the field decay, a plasma density step-up procedure is introduced to compensate the phase change and eventually a stable and nearly constant field is achieved. Recent study shows that in this case the acceleration process is almost linear \cite{Caldwell2}. Consequently, electron and positron beams can be accelerated up to 1 TeV in 2$\,$km long plasma, by using an LHC type proton beam. This plasma section, the beam focusing sections before the plasma, the beam delivery system and the instrumentation in the IP regions can be feasibly integrated into 4.3$\,$km LHC radius. Fig.\ref{fig:LHCILC} shows a schematic layout of a 2$\,$TeV CoM energy $e^+e^-$ collider located at the LHC tunnel, with the plasma accelerator cells marked in red. 

In this design, the proton extraction beam lines, located at both ends of a straight tunnel within LHC are needed to extract and guide the beam to the plasma cells. Before entering the plasma cells, the beam lines are designed to focus the proton beams so as to match the plasma focusing force. After focusing, protons are injected into preformed plasmas and excite the wakefields. We expect that after a few metres propagation in the plasma and together with a plasma density step-up, a full beam modulation is finally set up and constant wakefields are excited. Externally provided electron and positron beams are injected into the plasma with a correct phase (e.g. via tuning the positions and angles of both beams, etc.) in order to be accelerated in the wakefields. After 2$\,$km in plasma, a 1$\,$TeV electron beam and positron beam can be produced assuming an average accelerating field of 0.5 GeV/m in the plasma which is quite modest in comparison with the nominal achievable gradients by PDPWA technology \cite{Caldwell2}. A 2$\,$km beam delivery system for both electron and positron beams will transport and focus them to the IP, which is located in the middle of the tunnel, for collisions. After interacting with the plasma, the proton bunches will be extracted and dumped. These spent protons may also be recycled by the cutting-edge technologies, e.g. FFAG-based energy recovery \cite{Yakimenko} for reuse as driver beam or used to trigger the nuclear power plants \cite{Seryi}.

For this PDPWA-based $e^+e^-$ collider design, half of the LHC bunches (1404 bunches) are used for driving electron acceleration and the other half for positron acceleration. Taking into account the ramping time of the LHC is about 20 minutes and assuming the loaded electron (and positron) beams have a bunch charge of 10\% of the drive proton bunch, i.e. electron (and positron) bunch charge of  $N_e = 1.15\times10^{10}$, and the beam spot sizes at IP are the same as that of the CLIC beam, as shown in Table 1, the resulting luminosity for such an $e^+e^-$ linear collider is about $3.0\times10^{31} cm^{-2}s^{-1}$, which is about 3 orders of magnitude lower than that of the ILC or the CLIC.

\section{An electron-proton collider}
\label{}

One could also envisage an $ep$ collider design based on PDPWA scheme utilizing the CERN accelerator complex. The advantage of this design is based on the fact that the plasma-based option may be more compact and cheaper since it does not need to build an expensive and conventional 60 GeV electron accelerator, as proposed at the current LHeC design \cite{Bruening}.

In one of our designs, the SPS beam is used as the drive beam for plasma wakefield excitation. The reason for that is due to the long LHC beam ramping time (20 minutes). During the LHC beam energy ramping up from 450$\,$GeV to 7$\,$TeV, the SPS can prepare the drive beams (ramping time of LHC preinjectors is about 20 seconds) and then excite the wakefields and accelerate an externally injected low energy (e.g., tens of MeV) electron beam. When the accelerated electron beam is ready, it can be delivered to the collision points in the LHC tunnel for electron-proton collision. PIC simulation shows that working in the self-modulation regime, a wakefield amplitude of 1$\,$GeV/m can be achieved by using the SPS beam at an optimum condition where both the beam and plasma parameters are optimized \cite{Xia3}. Similar to the $e^+e^-$ collider design, the SPS beam needs to be guided to the plasma cell. Prior to the plasma cell, a focusing beam line is needed to match the beam with the plasma beta function. A $\sim$170$\,$m plasma cell is used to accelerate the electron beam up to 100 GeV. The energetic electrons are then extracted to collide with the circulating 7$\,$TeV proton beam. This parasitic $ep$ collision mode should allow LHC proton-proton collisions to continue in parallel.

The CoM energy in this case is given by,

\begin{equation}
\sqrt{s}=2\sqrt{E_eE_p} = 1.67\,TeV
\label{eqn:Luminosity2}
\end{equation}
                                      
where it is about a factor of 1.3 higher than the current LHeC design and a factor of 5.5 higher than the late HERA \cite{Diaconu}.
    The luminosity of an $ep$ collider for round and transversely matched beams is given by \cite{Wiik},
                                      
\begin{equation}
\mathcal{L}_{ep}=\frac{1}{4\pi}\frac{P_e}{E_e}\frac{N_p}{\epsilon^N_p}\frac{\gamma_p}{\beta^*_p},
\label{eqn:Luminosity2}
\end{equation}                                                                                               
where $P_e$ is the electron beam power, $E_e$ is the electron beam energy, $N_p$ is the number of particles in the proton bunch, $\epsilon_p^N$ is the normalized emittance of the proton beam, $\gamma_p$ is the Lorentz factor and $\beta_p^*$ is the beta function of the proton beam at the interaction point. The electron beam power is given by,
                                                                                               
\begin{equation}
P_e = N_eE_en_bf_{rep},
\label{eqn:Luminosity2}
\end{equation}
where $N_e$  is the number of particles in the electron bunch, $n_b$ is the number of bunches in the linac pulse and $f_{rep}$ is the repetition rate of the linac. Using the LHC beam parameters, for example, $N_p = 1.15\times10^{11}$, $\gamma_p = 7460$, $\beta^*_p = 0.1\,$m,   $\epsilon^N_p =3.5\, \mu$m and assuming the electron beam parameters as follows: $Ne = 1.15\times10^{10}$ ($10\%$ of the loaded drive bunch charge), $E_e$ = 100 GeV, $n_b = 288$ and $f_{rep}  \approx 15$, the calculated luminosity of the electron proton collider is about $1\times 10^{30}\,cm^{-2}s^{-1}$ for this design, which is about 3-4 orders of magnitude lower than the current LHeC design of $10^{33}$ or even $10^{34} cm^{-2}s^{-1}$. However, if one can increase the electron bunch intensity and the repetition rate, it may be possible to get a higher luminosity $ep$ collider based at CERN accelerator complex.
                                                                                                                                                      
\section{Some key issues in collider design}
\label{}                                                                                                                                                      
                                                                                                                                                      
\subsection{Phase slippage}
\label{}                                                                                                                                                        

Surfing on the right phase of the plasma wakefields driven by high-energy proton bunches, the electrons can be quickly accelerated to the relativistic energy regime. Due to the heavy mass of protons, the relativistic factor $\gamma$ of a TeV proton beam is smaller than that of an electron beam with energy of $1\,GeV$. Therefore the electrons may overrun the wakefields (the group velocity of the wakefields is the same as the velocity of the driver) and the acceleration process will be terminated. This phase slippage (dephasing) effect therefore becomes a limiting factor for a PDPWA-based collider, especially when a single plasma acceleration length extends over kilo meters.

We estimate that in the following conditions the significant dephasing can be avoided in a PDPWA-based collider design. To simplify the problem, we assume the wakefield structure in the co-moving frame does not evolve in time. It means that the protons (electrons) experience a constant deceleration (acceleration) field of magnitude $E_{dec}$ ($E_{acc}$). The rate of change of proton (with charge q) and electron (with charge e) energy are written as
    
\begin{equation}
\frac{d(\gamma_i m_i c^2)}{dt} = -qE_{dec}\nu_i,
\label{eqn:label}
\end{equation}    

\begin{equation}
\frac{d(\gamma_e m_e c^2)}{dt} = eE_{acc}\nu_e,
\label{eqn:label}
\end{equation}    
where $\gamma_i$, $m_i$ and $\nu_i$ are the relativistic gamma factor, mass and velocity of proton, respectively. $\gamma_e$, $m_e$ and $\nu_e$ are the relativistic gamma factor, mass and velocity of electron, respectively,  and c is the speed of light.

The relative position change between an electron and a proton at a time $T$ is given by \cite{Ruth}

\begin{equation}
\Delta s = \int^T_0(\nu_e-\nu_i)dt = \frac{m_e c^2}{e}\Big[ \frac{\gamma_{ef}-\gamma_{e0}}{E_{acc}} + \frac{m_ie}{m_eq}\frac{\gamma_{if}-\gamma_{i0}}{E_{dec}} \Big],
\label{eqn:label}
\end{equation}
where $\gamma_{e0}$, $\gamma_{ef}$ are the relativistic factor of the initial and final electron energies,  $\gamma_{i0}$, $\gamma_{if}$ are the relativistic factor of the initial and final proton energies, respectively.

The equations for the momentum are

\begin{equation}
\frac{d(\gamma_i m_i \nu_i)}{dt} = -qE_{dec},
\label{eqn:label}
\end{equation}

\begin{equation}
\frac{d(\gamma_e m_e \nu_e)}{dt} = eE_{acc}.
\label{eqn:label}
\end{equation}

Integrating the above momentum equations from 0 to $T$ gives

\begin{equation}
m_i c \Big(\sqrt{\gamma^2_{if}-1} - \sqrt{\gamma^2_{i0}-1} \Big) = -qE_{dec}T,
\label{eqn:label}
\end{equation}

\begin{equation}
m_e c \Big(\sqrt{\gamma^2_{ef}-1} - \sqrt{\gamma^2_{e0}-1} \Big) = eE_{acc}T,
\label{eqn:label}
\end{equation}

Combining the above two equations, we have

\begin{equation}
\Delta s = \frac{m_e c^2}{eE_{acc}} \Big( \gamma_{ef} - \gamma_{e0} \Big) 
\Bigg[   1 - \frac{\Big(\sqrt{\gamma^2_{ef}-1} - \sqrt{\gamma^2_{e0}-1} \Big) (\gamma_{if}-\gamma_{i0}) }
{\Big(\sqrt{\gamma^2_{if}-1} - \sqrt{\gamma^2_{i0}-1} \Big)(\gamma_{ef}-\gamma_{e0})}  \Bigg].
\label{eqn:label}
\end{equation}

It is worth noting that the relative position depends on the plasma density implicitly through the accelerating field $E_{acc}$. It also depends on the initial and final energies of the proton and electron. For the case $\gamma_{ef} \gg \gamma_{e0} \gg1$, the above equation can be written as

\begin{equation}
\Delta s \approx \frac{m_e c^2}{eE_{acc}} \Big( \gamma_{ef} - \gamma_{e0} \Big) 
\Bigg[   1 - \frac{(\gamma_{if}-\gamma_{i0}) }
{\Big(\sqrt{\gamma^2_{if}-1} - \sqrt{\gamma^2_{i0}-1} \Big)}  \Bigg]
\label{eqn:label}
\end{equation}

We can rewrite it in a phase slippage as

\begin{equation}
\delta = k_p \Delta s \approx \frac{1}{eE_{acc}/m_e c \omega_p} 
\Big( \gamma_{ef} - \gamma_{e0} \Big) 
\Bigg[   1 - \frac{(\gamma_{if}-\gamma_{i0}) }
{\Big(\sqrt{\gamma^2_{if}-1} - \sqrt{\gamma^2_{i0}-1} \Big)}  \Bigg],
\label{eqn:label}
\end{equation}
where $k_p = \omega_p / c $ is the plasma wave number, $\omega_p = (n_pe^2 / \epsilon_0m_e)^{1/2}$ is the plasma electron frequency, $n_p$ and $\epsilon_0$ are the plasma density and the permittivity of free space, respectively. To avoid phase slippage over acceleration length L, $\delta$ must be less than $\pi$, otherwise the electrons will overrun the protons.

For a single stage PDPWA based $e^+e^-$ collider design, a 7$\,$TeV LHC proton beam will excite plasma wakefields and accelerate electron bunches to 1$\,$TeV (assuming electron injection energy of 10$\,$GeV which is far less than 1$\,$TeV), $\gamma_{i0} \approx 7000$, $\gamma_{ef} - \gamma_{e0} \approx 2\times10^6$. If we assume that the amplitude of wakefields is $eE_{acc}/m_ec\omega_p \sim1$, then the phase slippage is 

\begin{equation}
k_p \Delta s  = 2\times10^6 \Bigg[ 1- \Big( \gamma_{if} - 7000  \Big) /  \Bigg(  \sqrt{\gamma^2_{if} - 1} - \sqrt{7000^2-1} \Bigg) \Bigg].
\label{eqn:label}
\end{equation}

The calculation shows that the phase slippage length (or maximum acceleration length) is about $\sim4\,$km assuming the plasma density of $10^{15}\,cm^{-3}$ for a final proton beam energy of around 1$\,$TeV. Therefore a 2$\,$km acceleration channel meets the phase slippage requirement for an $e^+e^-$ collider design.

\begin{figure}[ht] 
\centering
\includegraphics[width=0.5\textwidth]{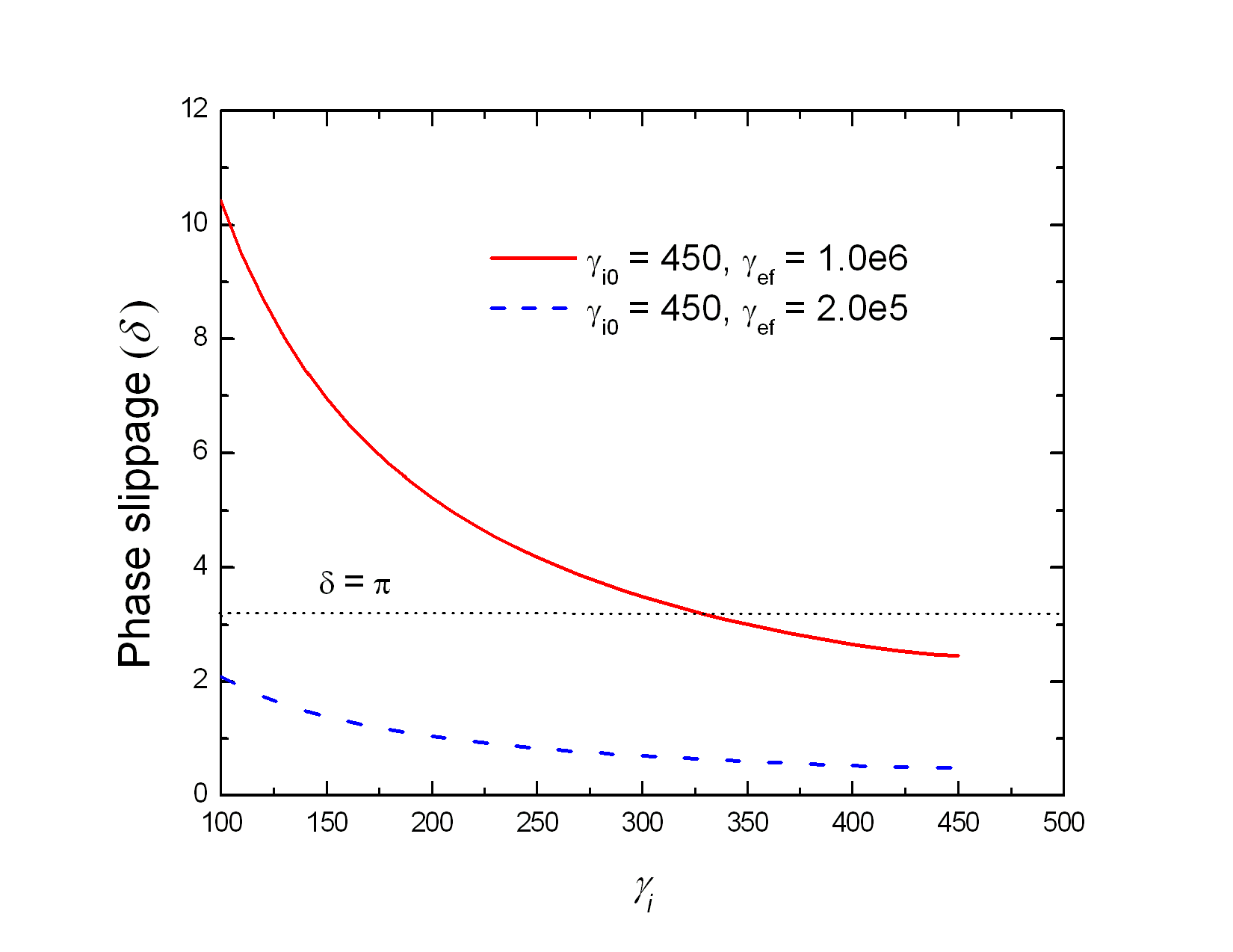}
\caption{Phase slippage between the SPS proton beam and the electron beam as a function of $\gamma_i$  of the proton drive beam for a single $500\,GeV$ stage and $100\,GeV$ stage electron beam production.}
\label{fig:SPS_phaseslippage}
\end{figure}

Since the SPS beam energy is much lower than the 7$\,$TeV LHC beam, phase slippage may become a problem if it is used as drive beam in a PDPWA-based collider design. Here, we consider two cases, one is to use the SPS beam to accelerate the electron beam up to 500$\,$GeV and the other one considers acceleration to 100$\,$GeV. The phase slippage for both cases are shown in Fig. \ref{fig:SPS_phaseslippage}.  For 500$\,$GeV electron acceleration, the final energy of the proton beam should be larger than 330$\,$GeV so as to satisfy the phase slippage requirement. If we use the average accelerating (decelerating) field of $\sim1\,$GeV/m at a plasma density of $10^{15}\,cm^{-3}$, the maximum dephasing length is about 170$\,$m. This provides the basic parameter to design such an acceleration stage. For 100$\,$GeV electron beam production, the phase slippage is always in the safe region. Therefore for a SPS drive beam, producing a 100$\,$GeV beam seems reasonable. 

\subsection{Proton propagation in the plasma}
\label{}       

To accelerate electrons (or positrons) to TeV energies, the acceleration length of a plasma cell needs to be of the order of several hundred or a few thousand meters, assuming an average accelerating gradient of $\sim$1$\,$GeV/m. In this case, the drive beam needs to propagate stably in such a long plasma cell without significant spreading. In vacuum, the beta function of the beam is $\beta_b = \beta\gamma\sigma_r^2/\epsilon_n$, where $\beta$ and $\gamma$ are the relativistic factors of the drive beam and $\sigma_r$ and $\epsilon_n$ are the $rms$ size and the normalized emittance of the drive beam, respectively. Taking the LHC beam as an example, $\beta \approx 1$, $\gamma \approx 7000$, $\sigma_r = 100\,\mu$m, $\epsilon_n = 3.5\,$mm.mrad, one has $\beta_b=20\,$m, which is far less than the required acceleration length. Therefore it is clear that some sort of transverse focusing is required in order to guide the drive beam over such a long distance. In principle, the transverse focusing can be external, e.g. by quadrupole magnets \cite{Caldwell1} or from the focusing force due to the transverse plasma wakefields. On the other hand, when the proton bunch propagates in the plasma, its finite momentum spread will induce a lengthening of the bunch. This can be evaluated for vacuum propagation as follows:

\begin{equation}
\Delta d \approx \frac{L}{2\Delta \gamma^2} \approx \frac{\Delta p}{p} \frac{m_p^2 c^4}{p^2 c^2} L
\label{eqn:label}
\end{equation}
where $\Delta d$ is the spatial spread of the particles in the bunch induced by the finite momentum spread $\Delta p/p$, $L$ is the distance travelled in the vacuum, $m_p$ is the proton mass, $p$ is the proton momentum and $c$ is the speed of light. For a $7\,$TeV LHC proton beam, $\Delta p/p = 10^{-4}$, the momentum spread leads to a growth of about $0.01\,\mu m\,m^{-1}$, which is negligible. Therefore large relative momentum spreads will still allow for long plasma-acceleration stages provided the drive beam is ultra relativistic.

\subsection{Electron-plasma interations}
\label{}     

For any above-mentioned TeV class collider design, the length of the plasma source is $\sim$km. One may have to consider electron scattering effects inside the long plasma cell.

An electron beam travelling through the plasma channel might undergo elastic and inelastic interactions with the plasma ions and plasma electrons with interaction cross sections depending on the beam energy and the characteristics of the plasma. In this section, the elastic scattering between the beam electrons and the plasma ions is investigated in pursuit of the resulting emittance growth in the electron beam. Assuming the plasma ions are stationary compared to the relativistic electrons, electrons are deflected by the nuclei via Coulomb scattering with the below scattering cross section,

\begin{equation}
\frac{d\sigma}{d\Omega}\approx(\frac{2Zr_{0}}{\gamma})^{2}\frac{1}{(\theta^{2}+\theta_{min}^{2})^{2}}
\label{eqn:label}
\end{equation} 
where $Z$ is the atomic number, $r_{0}$ is the classical electron radius, $\theta$ is the scattering angle, and $\theta_{min}\approx\hbar/pa$, where $a$ is the atomic radius given by $a\approx1.4\hbar^{2}/m_{e}e^{2}Z^{1/3}$, and $p$ is the incident particle momentum. 

The emittance growth caused by the elastic interaction of the electron beam and the plasma ions can be derived considering the previous work on beam-gas scattering in a damping ring \cite{TOR}. Therefore the emittance evolution of the electron beam inside the plasma cell can be written as the following, 

\begin{equation}
\gamma\epsilon_{x,y}(t)=\gamma(t)\frac{\tau}{2}\overline{\mathcal{N}\langle\theta_{x,y}^{2}\rangle\beta_{x,y}}
\label{eqn:emitt_1}
\end{equation}  

where $\mathcal{N}$ is the scattering rate, $\langle\theta^{2}\rangle$ is the expected value of $\theta^{2}$ and bar denotes the average along the plasma section. Simulations have shown that the energy of the electron beam linearly increases in the plasma channel as a function of time $t$ \cite{Caldwell2}. If $\gamma_{0}$ is the energy of the beam in the entrance of the plasma section, $g$ is the rate of change of $\gamma$. The following relation can be assumed for a beam accelerating linearly in the plasma channel:

\begin{equation}
\gamma(t)=gt+\gamma_{0}
\label{eqn:gamma_t}
\end{equation}  

For the time being, the damping term in the original approach will be modified by replacing the damping factor (emittance evolution in a damping ring $\epsilon_{y}(t)=\epsilon_{y}(0)exp(-2(t/\tau_{y}))$ where $\tau_{y}/2$ is time duration when the vertical emittance reduces down to a factor of $1/e$ of its initial value.) $(\tau_{y}/2)$ with $\tau$, the time duration that the beam travels in the plasma channel. $\mathcal{N}\langle \theta^{2}\rangle$ is given as Eq. (\ref{eqn:scatt_rate}) where $n_{gas}$ is the number density of the gas,

\begin{equation}
\mathcal{N}\langle\theta^{2}\rangle=cn_{gas}\int_{0}^{\theta_{max}}\frac{d\sigma}{d\Omega}\pi\theta^{3}d\theta.
\label{eqn:scatt_rate}
\end{equation}  

Consequently, the emittance evolution can be written as taking into account only the elastic scattering of the electrons by the nuclei in the plasma as given in Eq. (\ref{eqn:emitt_growth_plasmacell}) by substituting Eq. (\ref{eqn:gamma_t}) and Eq. (\ref{eqn:scatt_rate}) into Eq. (\ref{eqn:emitt_1}). 

\begin{equation*}
	 \Delta\epsilon_{n,\, scattering}(t)=(gt+\gamma_{0})
\end{equation*}  
\begin{equation*} 
 	\times \, \frac{\tau}{2}\langle cn_{gas}\beta\int_{0}^{\theta_{max}}(\frac{2Zr_{0}}{(gt+\gamma_{0}})^{2}\frac{1}{(\theta^{2}+\theta_{min}^{2})^	{2}}\pi\theta^{3}d\theta\rangle
\end{equation*}  

\begin{equation*}
=(gt+\gamma_{0})(\frac{2Zr_{0}}{(gt+\gamma_{0})})^{2}\frac{\tau}{2}\langle cn_{gas}\beta\int_{0}^{\theta_{max}}\frac{1}{(\theta^{2}+\theta_{min}^{2})^{2}}\pi\theta^{3}d\theta\rangle
\label{eqn:label}
\end{equation*}  
 
\begin{equation*} 
	=\frac{(2Zr_{0})^{2}}{gt+\gamma_{0}}\frac{\tau}{2}\langle cn_{gas}\beta\rangle\frac{\pi}{2\theta_{max}}
\end{equation*}   
\begin{equation}
	\times \, [3\theta_{min}tan^{-1}(\frac{\theta_{max}}{\theta_{min}})+\theta_{max}(log(\theta_{min}^{2}+\theta_{max}^{2})-2)]
\label{eqn:emitt_growth_plasmacell}
\end{equation}   

\begin{figure}[h] 
\centering
\includegraphics[width=0.5\textwidth]{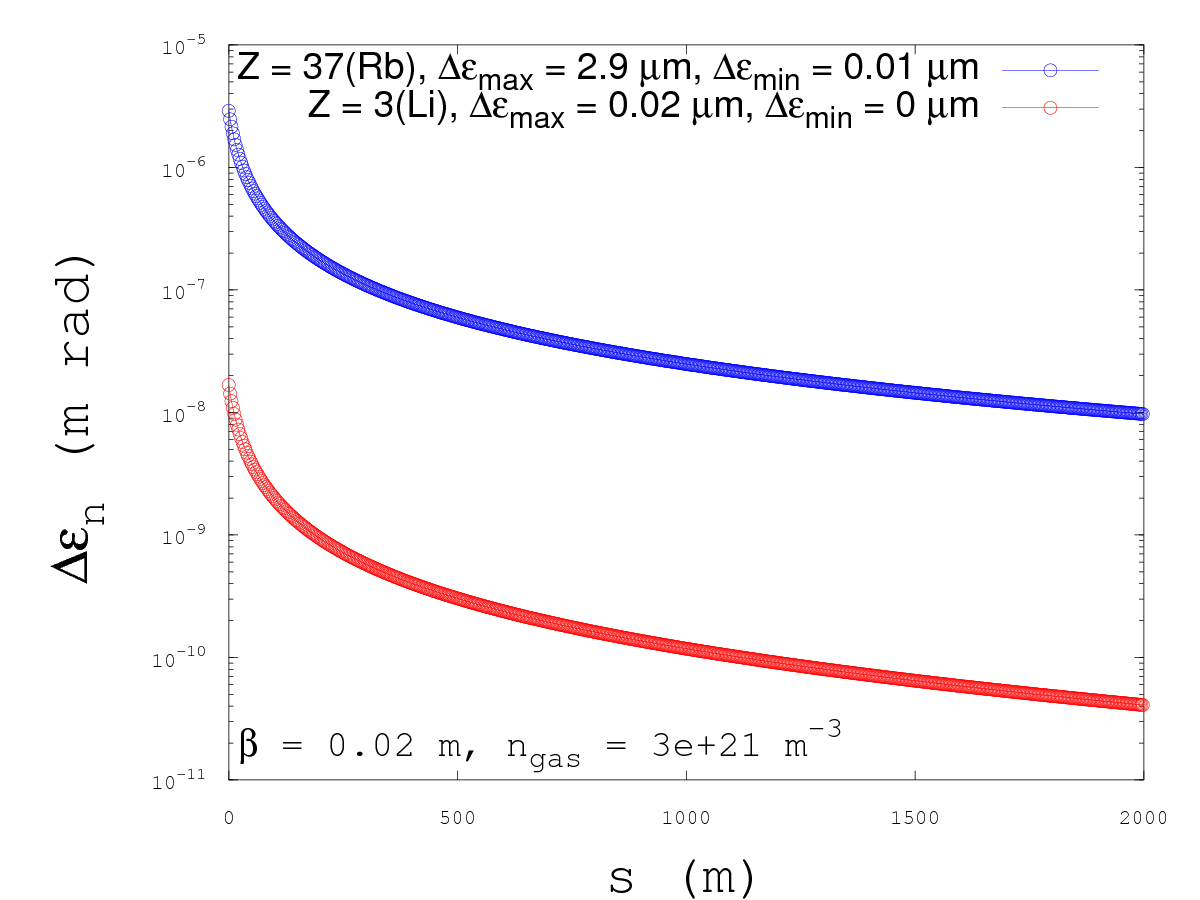}
\caption{The evolution of the emittance contribution from Coulomb scattering of the beam electrons by the plasma ions as a function of the distance travelled in a Rb ($Z$ = 37) and Li ($Z$ = 3) plasma.}
\label{fig:emitt_growth1}
\end{figure}

The evolution of the emittance contribution from the beam-nuclei scattering is shown in Fig. \ref{fig:emitt_growth1} as a function of the distance travelled in the plasma in the presence of different plasma forming gasses. Regardless of the element under consideration, the emittance growth falls rapidly with the linearly increasing energy through the plasma channel. In this study, the initial energy of the electron beam at the entrance of the plasma section is 10$\,$GeV. The emittance contribution from scattering with the Rb ($Z$ = 37) nuclei is $3\,\mu$m at this initial stage. Whereas, it decreases down to $0.01\,\mu$m in the exit of the plasma section where the beam is accelerated up to an energy of 1$\,$TeV. The contribution to the emittance is shown to be two orders of magnitude lower in the case of a lower-$Z$ element, Li ($Z$ = 3). The total emittance, at any time during the plasma acceleration, can be calculated through a quadratic sum of the contribution due to scattering and the design emittance, as shown in Eq. (\ref{eqn:scatt_contribution}).

\begin{equation} 
\epsilon_{n,\, total}=\sqrt{\epsilon_{n,\, design}^{2}+\Delta\epsilon_{n,\, scattering}^{2}}
\label{eqn:scatt_contribution}
\end{equation}   

The beam-plasma interaction is under further investigation in order to quantify the energy loss and the energy spread of the witness beam through the elastic scattering with the plasma electrons and the inelastic scattering with both plasma electrons and ions.

\subsection{Positron acceleration in PDPWA}
\label{}

Simulations have shown that a bunch of electrons can be accelerated by either a compressed proton-driven plasma wakefield acceleration scheme \cite{Caldwell1} or by a long proton bunch driven wakefield in self-modulation regime \cite{Caldwell2}. However, for any $e^+e^-$ linear collider design, a high-energy positron beam is also required for beam collision. The positron acceleration still needs to be investigated in more detail.

More recently a new scheme for accelerating positively charged particles in a plasma-wakefield accelerator has been proposed by Yi et al \cite{Yi}. In this scheme, the proton drive bunch propagates in a hollow plasma channel, and the channel radius is of the order of the beam radius. The space charge force of the driver beam causes charge separation at the channel wall, which helps to focus the positively-charged witness bunch propagating along the beam axis. In the plasma channel, the acceleration buckets for positively charged particles are much larger than in the blowout regime of the uniform plasma, and a stable acceleration over a long distance is possible. In addition, the phasing of the witness with respect to the wave can be tuned by changing the radius of the channel to ensure the optimal acceleration. The performed two-dimensional simulations have shown that a 2$\,$TeV LHC-like beam, longitudinally compressed to 100 $\mu$m, with a bunch intensity $10^{11}$ and energy spread $10\%$ can excite a strong wakefield and accelerate a witness $2\,$TeV proton bunch with bunch charge of $1\,$nC, injected at 0.75 mm behind the drive beam, over 1$\,$km in a hollow plasma channel with the plasma density of $6 \times 10^{14}\,cm^{-3}$. The resulting energy gain for the witness proton beam is over 1.3$\,$TeV in a 1$\,$km plasma channel. 

At high energies, protons behave very similarly to positrons; positrons can certainly be accelerated with this scheme. The detailed 3D PIC simulations are now underway to verify the positron acceleration effect in a hollow plasma channel.

\section{Other novel ideas}
\label{}

Many novel ideas have emerged since the PDPWA concept has been proposed. Recent simulations have shown that a $10 \sim 100\,$GeV proton bunch with a bunch length less than $100\,\mu$m can be generated with a laser intensity of $10^{22}\,W/cm^2$ via a so-called snowplow regime of the laser-driven wakefield acceleration \cite{Zheng}. One may think of injecting such a short and high-energy proton bunch into a fast cycling synchrotron to boost the beam energy quickly (up to $\sim$TeV) while keeping the short proton bunch length. This resulting high energy, short proton bunch may be used as an ideal driver to resonantly excite a large amplitude plasma wakefield for electron beam acceleration and for a collider design based on the PDPWA scheme. This method may also serve as a preparation for TeV regime acceleration of protons over centimeters with a laser pulse with peak power of $10^{23}\, W/cm^2$, e.g. a laser from the Extreme Light Infrastructure-ELI which is under construction \cite{link}.
    
Seryi proposed a multi-TeV upgrade concept for the ILC based on PDPWA scheme \cite{Seryi}. In this concept the proton bunches are accelerated together with electrons and positrons simultaneously by employing the ILC technology (1.3$\,$GHz superconducting RFs). A special beamline arrangement would allow control of proton phase slippage, separation and merging of proton and electron (positron) bunches via dual-path chicanes, as well as ballistic compression of the proton bunches. This approach may open a path for the ILC to a much higher energy upgrade to several TeVs. 
 
Yakimenko et al also discussed a possible solution to a TeV CoM $e^+e^-$ linear collider design based on PDPWA concept. Such an $e^+e^-$ collider may use the proton beam from the Tevatron as driver and fit into a 6.3$\,$km tunnel. In this scheme, a high average power proton drive beam is required for exciting the plasma wakfields for electron and positron beam acceleration. The spent proton beams (with significant amount of energy) will be recycled for further energy boost to 1$\,$TeV by the FFAG fast cycling rings \cite{Yakimenko}. This scheme may be able to increase the collision repetition rate and therefore the collider luminosity significantly.

\section{Conclusions}
\label{}

Simulations have shown that either a longitudinally compressed (e.g. $100\, \mu$m) or an uncompressed long proton bunch can be used to drive a large amplitude plasma wakefields and accelerate an electron beam to the energy frontier in a single stage. We therefore conceive of an $e^+e^-$ collider and an $ep$ collider design based on this scheme.  Using the LHC beam as the drive beam, it is possible to design a 2$\,$TeV CoM energy $e^+e^-$ collider in the LHC tunnel. For an $ep$ collider design, the SPS beam can be used as the drive beam to accelerate an electron beam up to $\sim100\,$GeV. The CoM energy in this case is 1.67$\,$TeV, which is greater than that of the current LHeC design. It is worth noting that although the luminosity is not as high as that of the ILC, CLIC or the LHeC (about two to three orders magnitude lower), there are still many interesting physics which can be addressed by using very high energy but low luminosity $e^+e^-$ collider or $ep$ collider, such as classicalization in electroweak processes, study of QCD and beyond standard model physics and study of source of high energy cosmic rays, etc \cite{Bartels}.  For a TeV linear collider design, phase slippage between the proton beam and electron (positron) beam may become a limiting factor for $\sim$km plasma accelerator. 

\section*{Acknowledgment}
\label{}

This work is supported by the Cockcroft Institute Core Grant and STFC.

%\section*{References}


%% References


%% \bibliographystyle{elsarticle-num}
%% \bibliography{<your-bib-database>}


\bibliographystyle{unsrt}

\end{document}